\documentclass[journal,onecolumn]{IEEEtran}
\usepackage{cite}
\usepackage{algorithmic}
\usepackage{textcomp}
\usepackage{float}
\usepackage{multirow}
\usepackage{amssymb,amsmath,amsthm,enumitem}
\usepackage{mathtools}
\usepackage{subfig}
\usepackage{amsfonts}
\usepackage{graphicx}
\usepackage{adjustbox}
\usepackage[T1]{fontenc}
\usepackage{array}
    \newcommand{\RNum}[1]{\uppercase\expandafter{\romannumeral #1\relax}}
\usepackage{mathtools}
\usepackage{array}
\newcolumntype{P}[1]{>{\centering\arraybackslash}p{#1}}
\newcolumntype{M}[1]{>{\centering\arraybackslash}m{#1}}

    \newcommand*{\matminus}{%
  \leavevmode
  \hphantom{0}%
  \llap{%
    \settowidth{\dimen0 }{$0$}%
    \resizebox{1.1\dimen0 }{\height}{$-$}%
  }%
}

\newcommand*{\matplus}{%
  \leavevmode
  \hphantom{0}%
  \llap{%
    \settowidth{\dimen0 }{$0$}%
    \resizebox{1.1\dimen0 }{\height}{$+$}%
  }%
  }

\begin{document}

\title{Ramanujan Periodic Subspace Division Multiplexing(RPSDM)\\
\thanks{Identify applicable funding agency here. If none, delete this.}
}

\author{\IEEEauthorblockN{Goli Srikanth\IEEEauthorrefmark{1},
Vijay Kumar Chakka\IEEEauthorrefmark{2} and Shaik Basheeruddin Shah\IEEEauthorrefmark{3}}\\
\IEEEauthorblockA{Department of Electrical Engineering, \\
Shiv Nadar University, Greater Noida, Uttar Pradesh-203207\\
Email: \IEEEauthorrefmark{1}gs499@snu.edu.in,
\IEEEauthorrefmark{2}bs600@snu.edu.in,
\IEEEauthorrefmark{3}Vijay.Chakka@snu.edu.in}}

\maketitle

\begin{abstract}
In this paper, a new modulation method defined as Ramanujan Periodic Subspace Division Multiplexing (RPSDM) is proposed using Ramanujan subspaces. Each subspace contains an integer valued Ramanujan Sum (RS) and its circular downshifts as a basis. 
The proposed RPSDM decomposes the linear time-invariant wireless channels into a Toeplitz stair block diagonal matrices, whereas Orthogonal Frequency Division Multiplexing (OFDM) decompose the same into diagonal. Advantages of such structured subspaces representation are studied and compared with an OFDM representation in terms of Peak-Average Power Ratio (PAPR) and Bit-Error-Rate (BER). Zero Forcing (ZF) and Minimum Mean Square Error (MMSE) detectors are applied to evaluate the performance of OFDM and RPSDM techniques. Finally, the simulation results show that the proposed design (with an additional receiver complexity) outperforms OFDM under both detectors.\end{abstract}

\section{Introduction}
OFDM is one of the best solutions which fulfilled the requirements of the fourth generation (4G) networks \cite{4607239}. 
It is a well-known technique, which offers several advantages such as resiliency towards multipath fading \cite{54342} results in a one-tap equalizer in the frequency domain \cite{841722}. 
Despite the advantages, it suffers from high PAPR, out-of-band emissions and synchronization issues. 
Due to high PAPR, it requires a high dynamic range of the power amplifier leads to an increase in the effective cost of the system \cite{cho2010mimo}. To address these problems, a modified version of existing multicarrier modulation or an alternative transform based multicarrier method has to be investigated.\\

A modified version of an existing OFDM multicarrier modulation like  Filter Bank Multicarrier (FBMC) \cite{5753092}, \cite{farhang2014filter}, Universal Filtered Multicarrier (UFMC) \cite{6824990} and Generalized Frequency Division Multiplexing (GFDM) \cite{5073571} are proposed as emerging candidates for 5G modulation schemes \cite{6824752}. FBMC and UFMC are filtered version of the OFDM, which reduces the out of band emissions. FBMC concentrates on per subcarrier filtering and UFMC is based on subband filtering. Unlike filtering, GFDM is one such multicarrier modulation which provides specific subcarrier allocation, results in low PAPR and out-of-band emissions.  
On the other hand, different transform based representations apart from Discrete Fourier Transform (DFT) like Discrete Cosine Transform (DCT) \cite{482113}, Discrete Sine Transform (DST) \cite{1304952}, Discrete Hartley Transform (DHT) \cite{860954} and Discrete Wavelet Transform (DWT)  \cite{923351} have been proposed as multi-subspace representation methods. 
Mandyam \cite{1304952} and Sanchez \emph{et al.} \cite{482113} studied the assets of the real based sinusoidal transform for multicarrier systems.  Liang \emph{et al.} proposed DHT (sum of sine and cosine) \cite{5487429} which reduces the computational complexity compared to DFT. 
To probe the time-overlap benefits of communication channel a transform based mutltisubspace method like wavelet packet transform is proposed by Bouvel \emph{et al.} \cite{923351}. Due to the different time and frequency resolutions, the frequency behavior of wavelet transform is not straightforward. \\

Recently, P. P. Vaidyanathan proposed a finite length signal representation using linear combination of signals that belong to Ramanujan subspaces. Each subspace is spanned by Ramanujan Sum (RS) and its circular downshifts \cite{6839014}. RSs are an integer-valued sequences, introduced by the mathematician Srinivasa Ramanujan \cite{ramanujan1918certain} to represent well-known arithmetic functions.  In this paper, an alternate transform based modulation is proposed using the basis of Ramanujan periodic subspaces known as Ramanujan Periodic Subspace Division Multiplexing (RPSDM). The main motivation behind this proposal is to interpret the periodic behaviour of communication channels. Major contributions of the paper are as follows: 
\begin{itemize}
\item A novel modulation method defined as RPSDM is proposed for frequency selective fading channels.
\item Decomposition of the circulant channels into Toeplitz stair block diagonal structure using RPSDM.
\item The superiority of proposed system over an existing OFDM technique is validated through simulation results.

\item The worst case PAPR and computational complexity for RPSDM in comparison with an OFDM are derived.
\item Finally, merits and demerits of OFDM and RPSDM are tabulated.
\end{itemize}
 
The rest of the paper is organized as follows: Section \ref{systemmodel} describes the cyclic prefix based system model for frequency selective fading channel. Section \ref{method-techniques} presents the multicarrier modulation schemes of OFDM and RPSDM along with ZF and MMSE detectors. Section \ref{PAPRanalysis} discuss the PAPR analysis of OFDM and RPSDM. Simulation results, computational complexity and their comparisons are presented in Section \ref{Results}. Finally, we draw conclusions and future directions in Section \ref{confut}.

\section{System Model}\label{systemmodel}
Consider a Single Input and Single Output (SISO) frequency selective fading channel with rich scattering environment. The input-output relation of such systems are modeled as tapped delay line of length $L $ \cite{savaux2014mmse}. It is represented as Finite Impulse Response (FIR) system in time domain. Then the received (output) signal $y[n]$ at the $n^{th}$ symbol is denoted as,
\begin{equation}
y[n] =\sum_{l=0}^{L-1}h_{l} x[n-l]+ w[n],
\label{sysmod1}
\end{equation}
where, $x[n]$ is an input symbol for the $n^{th}$ instant, $w[n] $ is the \emph{circularly symmetric complex} Additive White Gaussian Noise (AWGN) with $\mathcal{N}(0,\sigma^2)$, $L$ is the total number of multipaths and $h_{l}$ is $l^{th}$ path delay between the transmitter and receiver. Therefore the overall Channel Impulse Response (CIR)  $\bf{h}$ of length $L$  multipaths are denoted as, 
\begin{equation}
\bold{h} = [\textit{h}_{0},\dots, \textit{h}_{L-1}]^{T},
\label{CIR}
\end{equation}
where, fading paths $\bold{h}$ are assumed to be complex \emph{Gaussian} distributed \emph{iid} random variables with $\mathcal{N}(0,1)$. \
\begin{figure}[ht]
\centering
\includegraphics[width=8cm,height=4cm]{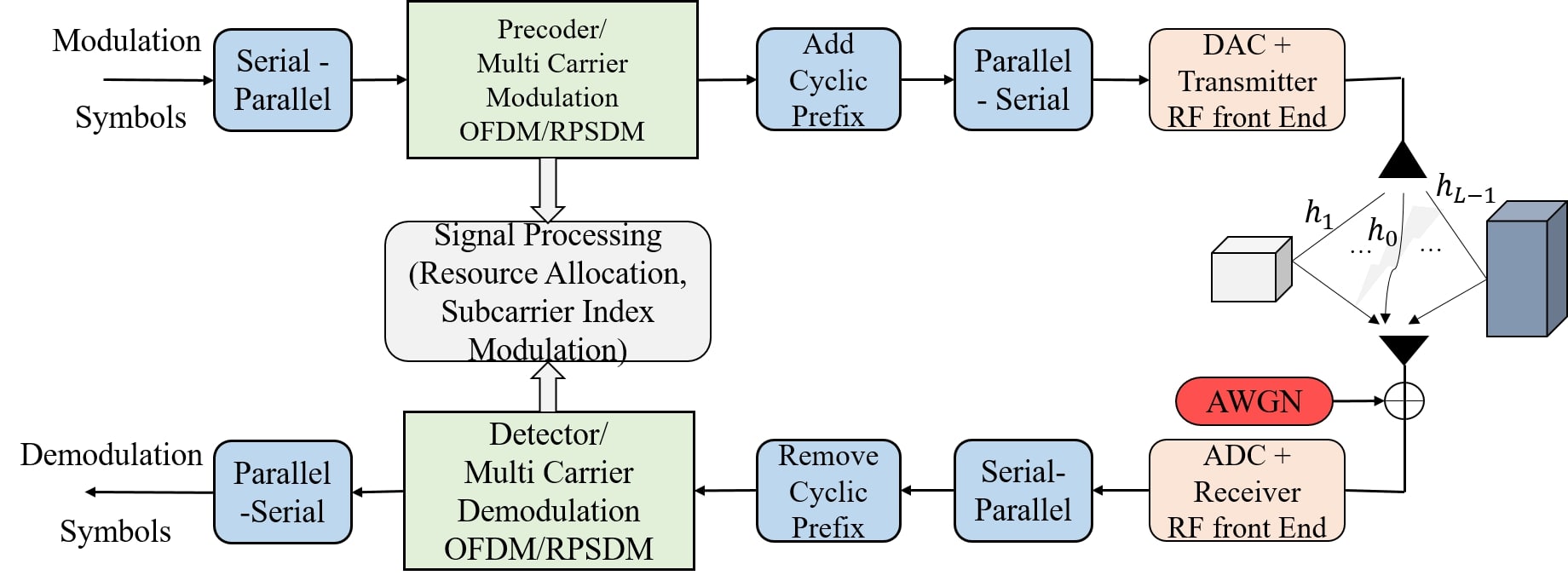}
\caption{\small{CP based Multicarrier Signal Processing Block Model}}
\label{blockdiagram}
\end{figure}

The transmitted vector $\bold{x}=  [x(0),x(1),\dots, x(N-1)]^{T}$ is converted in to frame $\bold{x}_{cp}$ using Cyclic Prefix (CP) based transmission as follows:
\begin{equation}
\bold{x}_{cp} = [x(N-L+1), x(N-L+2),\dots, x(N-1),\hspace{0.3cm} \bold{x}]^{T} .
\end{equation}

It occupies $K = N+L-1$ duration. 
Now, $K$ symbol block at the receiver converts the linear convolution of (\ref{sysmod1}) into circular convolution. Its equivalent model is represented as,

\begin{equation}
{\bold{\bar{y}}} =  \bold{\bar{H}}\bold{x}_{cp}+\bold{\bar{w}},
\label{sysmod3}
\end{equation}
where, $\bold{\bar{y}}$ is received vector, $\bold{\bar{w}}$ is noise vector and $\bold{\bar{H}}\in\mathbb{C}^{K\times K}$ is circulant matrix \cite{davis2012circulant}. The above system model \eqref{sysmod3} is transformed to \eqref{sysmod4} by CP removal ( i.e., $\bold{\bar{y}}\in \mathbb{C}^{K}$ becomes $\bold{y}\in \mathbb{C}^{N}$),
\begin{equation}
{\bold{y}} =  \bold{H}_{cir}\bold{x}+\bold{w},
\label{sysmod4}
\end{equation}
here $\bold{w}\in \mathbb{C}^{N}$ is the AWGN vector,  $\bold{H}_{cir}\in \mathbb{C}^{N\times N}$ is the circulant matrix and $\bold{x}$ is the multicarrier modulated data. To design the multicarrier modulated data, we consider  linear transformation $\bold{E}_{t}\in \mathbb{C}^{N\times N}$  defined as modulation matrix,
\begin{equation}
\bold{x} =  \sqrt{\frac{P}{N}}\bold{E}_{t}\bold{X},
\label{modulation}
\end{equation}
where, $P$ is the total transmit power, $\bold{X}\in\mathbb{C}^{N}$ are the digital modulation symbols selected from $M$-QAM constellation. It consists of $\mathcal{A} = \{ (2n_1-1-\sqrt {M})+j(2n_2-1-\sqrt{M})\}$  where, $n_1,n_2 \in \{1,2,\dots,\sqrt{M}\}       $ \cite{zareian2007analytical}.
 $\bold{E}_r \in \mathbb{C}^{N\times N}$ is considered as demodulation matrix to demodulate the received signal in \eqref{sysmod4}.  Under these transformations the effective system is written as,
\begin{equation}
\bold{Y} =   \sqrt{\frac{P}{N}}\bold{H}\bold{X}+\bold{W},
\label{sysmod_MC}
\end{equation} 
where, $\bold{W}$ $=\bold{E_{r}}\bold{w}$, $\bold{Y}$ $=\bold{E_{r}}\bold{y}$ are the transformed noise and received  vectors of size $N$ respectively and $\bold{H}=\bold{E}_{r}\bold{H}_{cir}\bold{E}_{t}$ is the transformed channel matrix of size $N\times N$. Pictorial representation of physical layer multicarrier block model is shown in Fig. \ref{blockdiagram}. 

\section{OFDM and RPSDM based Transformations}\label{method-techniques}
 In this section, modulation matrix $\bf{E}_{t}$ and demodulation matrix  $\bf{E}_{r}$ discussed above are designed using complex exponential basis and basis of Ramanujan subspaces.

\subsection{OFDM Modulation \& Demodulation}
For the modulation matrix, $\bold{E}_{t} = \frac{1}{\sqrt{N}}\big[s_{N,0}[n],s_{N,1}[n],\dots, s_{N,k}[n],$ $\dots,s_{N,N-1}[n]\big]$, where $s_{N,k}[n]=e^{\frac{j2\pi kn}{N}}$ is the $k^{th}$ column vector of $\bold{E}_{t}$ having length $N$. Then the above described system model \eqref{modulation}-\eqref{sysmod_MC} converts to OFDM based multicarrier system. The synthesized time domain signal is written as,
\begin{equation}
\label{ofdmsynth}
x[n] = \frac{1}{\sqrt{N}}\sum_{k=0}^{N-1}X[k]s_{N,k}[n],\quad 0\leq n \leq N-1,\ n{\in}\mathbb{Z}.
\end{equation}

Demodulation matrix $\bold{E}_{r}=\bold{E}_{t}^{H}$  is known as Discrete Fourier Transform (DFT) matrix. 
Therefore channel $\bf{H}$ in \eqref{sysmod_MC} is modified to diagonal matrix $\bf{H}_{d}$ having the elements of $N-$point DFT coefficients of vector $\bf{h}$ defined in \eqref{CIR},
\begin{equation}
\bold{Y} =   \sqrt{\frac{P}{N}}\bold{H}_{d}\bold{X}+\bold{W},
\label{sysmod_MCO}
\end{equation}
here, $\bold{Y}$ and $\bold{W}$ are received and noise vector under the transformation of complex exponential basis.

\subsection{RPSDM Modulation and Demodulation}
System model described in \eqref{modulation}-\eqref{sysmod_MC} is converted to RPSDM based system with modulation matrix $\bold{E}_{t} = [\bold{S}_{q_1},\bold{S}_{q_2},\dots, \bold{S}_{q_m}]\in \mathbb{Z}^{N\times N}$, where $q_{1}$ to $q_{m}$ are all $m$ divisors of $N$.  Each $\bold{S}_{q_i}\in \mathbb{Z}^{N\times \phi(q_i)}$ is given by,
\begin{equation}\label{Sqmatrix}
\bold{S}_{q_i} = \big[\hat{c}_{q_i}[n],\hat{c}_{q_i}[((n-1))_{q_i}],\dots, \hat{c}_{q_i}[((n-\phi(q_{i})))_{q_i}]\big],
 \end{equation}
where $\hat{c}_{q_i}[n]$, is an $N$ length column vector obtained by repeating $q_i$ periodic sequence $c_{q_i}[n]$ by $\frac{N}{q_i}$ times. $((n-\phi(q_{i})))_{q_i}$ indicates $(n-\phi(q_{i}))$ modulo $q_i$. Here $\phi{(q_i)}$ is totient function \cite{hardy1979introduction}  represents the number of integers in $1\leq l \leq q_i$ satisfying  gcd (\emph{greatest common divisor}) $(l,q_i) = 1$. $\bold{S}_{q_i}$ can be visualized as a column wise periodic repetition of $\bold{C}_{q_i}\in \mathbb{Z}^{{q_i \times  \phi{(q_i)}}}$ by $\frac{N}{q_i}$ times, where
\begin{equation}\label{Cqmatrix}
\bold{C}_{q_i}=
\big[{c}_{q_i}[n],{c}_{q_i}[((n-1))_{q_i}], \dots, {c}_{q_i}[((n-\phi(q_{i})))_{q_i}]\big],
\end{equation}
where $\bold{C}_{q_i}$ is the basis for column space of matrix  $\bold{D}_{q_i}\in \mathbb{Z}^{{q_i \times  q_i}}$ \cite{6839030} known as Ramanujan subspace shown in \eqref{Dqmatrix} and $c_{q_i}[n]$ is Ramanujan Sum (RS),
\begin{equation}\label{Dqmatrix}
\bold{D}_{q_i}=\begin{bmatrix}
{c_{q_i}}(0)        & {c_{q_i}}(q_i-1) & \dots   & {c_{q_i}}(1)\\
{c_{q_i}}(1)        & {c_{q_i}}(0)        & \dots   & {c_{q_i}}(2)\\
\vdots                  & \vdots                   & \ddots & \vdots\\
{c_{q_i}}(q_i-1) & {c_{q_i}}(q_i-2)  & \dots   & {c_{q_i}}(0)  \\
\end{bmatrix}.
\end{equation}
  Since $\sum_{{q_i}|N}\phi{(q_i)} =N$ \cite{hardy1979introduction}, then its synthesis equation is

\begin{equation}
x(n) =\sum_{{q_i}|N}{x_{q_i}(n)},
\label{ramsEq}
\end{equation}
summation in (\ref{ramsEq}) is executed for those $q_i$ values which are divisors of $N$. The sequence $x_{q_i}(n)$ belongs to Ramanujan subspace $\bold{D}_{q_i}$ expressed as follows,

\begin{equation}
\label{subspaceeq}
\begin{split}
x_{q_i}(n) &= \frac{1}{\sqrt{N\phi{(q_i)}}}\sum_{l=0}^{\phi{(q_i)}-1}{X}[l+q_{i}-\phi(q_{i})]\hat{c}_{q_i}(n-l)\\
\end{split}
\end{equation}

Now, we discuss RSs and their properties. The RS is a linear combination of complex exponential sums having  period $q_i$ and frequencies $\{\frac{2\pi l}{q_i}|1\leq l \leq q_i, (l,q_i)=1 \}$. From here on, we drop the index subscript $i$ for simplicity and stated otherwise. For any given $q \geq1$, the summation is defined as  $c_q[n]$ \cite{ramanujan1918certain}.
\begin{equation}
\label{eq8}
c_q[n]= \sum_{\substack{k=1\\(k,q)=1}}^{q}e^{\frac{j2{\pi}kn}{q}},\quad \forall n\in\mathbb{Z}\ \text{and}\ q\in\mathbb{Z^+},
\end{equation}
For example $q={4}$, $c_4[n] = [2, 0, -2, 0]$. 
Few properties of RSs, which are useful for defining RPSDM modulation are tabulated in Table \ref{Properties}. These can be verified easily.

\begin{table}[ht]
\centering
\begin{tabular}{ | M{5.1em} |  M{6.1cm} | } 
\hline
\textbf{Properties} & $c_{q}[n]$ \\ 
\hline
Periodicity &  $c_{q}[n+q]=c_{q}[n]$ \\ 
\hline
Orthogonality &  $
\sum_{n=0}^{m-1}c_{q_{1}}[n]c_{q_{2}}[n]=
  \begin{cases} 
   q_{1}\phi(q_{1}) & \text{if } q_{1} = q_{2} \\
   0       & \text{if } q_{1}\neq q_{2}
  \end{cases}
$ $m =lcm(q_{1},q_{2})$\\ 
\hline
$q_{i}$ is prime &  $
c_{q_{i}}[n]=
  \begin{cases} 
   q_{i}-1& \text{if $q_{i}|n$} \\
   -1       & \text{otherwise}
  \end{cases}
$ \\ 
\hline
$q_{i} = p^{t}$, $t>1$ power of prime &  $
c_{p^{t}}[n]=
  \begin{cases} 
   0 & \text{if $p^{t-1}\not|n $ }\\
  -p^{t-1}        & \text{if $p^{t-1}|n$ but $p^{t}\not|n$ }\\
  p^{t-1}(p-1) & \text{if $p^{t}|n$ }
  \end{cases}
$ \\ 
\hline
Multiplicative &
$c_{q_{i}q_{j}}[n]=c_{q_{i}}[n]c_{q_{j}}[n]$  \text{for }  ($q_{i},q_{j}$)=1. \\ 
\hline
\end{tabular}
\caption{Properties  of Ramanujan Sum $c_{q}[n]$}
\label{Properties}
\end{table}

Here the notation $a\not|b$ indicates $a$ is not a divisor of $b$.
A different way to see orthogonality property mentioned in Table   \ref{Properties} is by non-overlapping DFT coefficients of RSs \cite{6839014}, which is discussed below.\\

\emph{Non-overlapping DFT Coefficients}: Let $c_{q_1}(n)$ and $c_{q_2}(n)$ are two different RSs having periods $q_1$ and $q_2$ respectively, and choose a positive integer $N$ such that $q_1|N$ and $q_2|N
$. Then the $N$-point DFT of $c_{q_1}(n)$ is
\begin{equation}
C_{q_1}[k] = \begin{cases}
N,\ &\text{if}\ k_1l_1=k,\ \text{where}\ l_1 =  N/{q_1}\\\ &\text{and}\ 1{\leq}k_1{\leq}q_1\ \textit{s.t.}\ (k_1,q_1)=1,\\
0\ &\ \text{Otherwise}.
\end{cases}
\end{equation}
similarly
\begin{equation}
C_{q_2}[k] = \begin{cases}
N,\ &\text{if}\ k_2l_2=k,\ \text{where}\ l_2 = N/{q_2}\\\ &\text{and}\ 1{\leq}k_2{\leq}q_2\ \textit{s.t.}\ (k_2,q_2)=1,\\
0\ &\ \text{Otherwise}.
\end{cases}
\end{equation}
If $k_1l_1 = k_2l_2$, this leads to $q_1 = q_2$, so DFT coefficients of two RSs never overlap. Using above discussion, the matrix representation of \eqref{ramsEq} for $N = 4$ is shown in \eqref{ramsmatrix},
\begin{equation}\label{ramsmatrix1}\nonumber
\begin{aligned}
x(n)  = \frac{1}{\sqrt{N\phi(q_1)}}X[0]{\hat{c}_1}[n]+\frac{1}{\sqrt{N\phi(q_2)}}X[1]{\hat{c}_2}[n]+ \\
\frac{1}{\sqrt{N\phi(q_4)}}X[2]{\hat{c}_4}[n]+\frac{1}{\sqrt{N\phi(q_4)}}X[3]{\hat{c}_4}[n-1],
\end{aligned}
\end{equation}

\begin{equation}\label{ramsmatrix}
\begin{aligned}
\bold{x} & = \mathbf{Q}\underbrace{\begin{bmatrix}
{c_1}(0) & {c_2}(0) & {c_4}(0) & {c_4}(3)\\
{c_1}(0) & {c_2}(1) & {c_4}(1) & {c_4}(0)\\
{c_1}(0) & {c_2}(0) & {c_4}(2) & {c_4}(1)\\
\underbrace{{c_1}(0)}_{\bold{S}_1} & \underbrace{{c_2}(1)}_{\bold{S}_2} & \underbrace{{c_4}(3)}_{\bold{S}_4} & \underbrace{{c_4}(2)}_{\bold{S}_4}
\end{bmatrix}}_{\bold{E}_t}
\underbrace{\begin{bmatrix}
X[0]\\
X[1]\\
X[2]\\
X[3]
\end{bmatrix}}_{\bold{X}},\\ & = \bold{Q}{\bold{E}_t}{\bold{X}},
\end{aligned}
\end{equation}

where $\bold{Q}$ is the diagonal matrix with elements having respective subspace normalization factors as shown below:
\footnotesize\begin{equation}                                                                                                                                                                                                                                                                                                                                                                                                                                                                                                                                                                                                                                                                                                                                                                                                                                                                                                                                                                                                                                                                                                                                                                                                                                                                                                                                                                                                                                                                                                                                                                                                                                                                                                                                                                                                                                                                                                                                                                                                                                                                                                                                                                                                                                                                                                                                                                                                                                                                                                                                                                                                                                                                                                                                                                                                                                                                                                                                                                                                                                                                                                                                                                                                                                                                                                                                                                                                                                                                                                                                                                                                                                                                                                                                                                                                                                                                                                                                                                                                                                                                                                                                                                                                                                                 \bold{Q} =\begin{bmatrix}
    \frac{1}{\sqrt{N\phi(q_1)}}\bold{I}_{\phi(q_1)}& \bold{0}_{\phi(q_1)\times \phi(q_2)} &\dots&\bold{0}_{\phi(q_1)\times \phi(q_m)}\\
    \bold{0}_{\phi(q_2)\times \phi(q_1)}&\frac{1}{\sqrt{N\phi(q_2)}}\bold{I}_{\phi(q_2)}  & \dots&\bold{0}_{\phi(q_2)\times \phi(q_m)}\\
     \vdots &\vdots & \ddots &\vdots\\
    \bold{0}_{\phi(q_m)\times \phi(q_1)} & \bold{0}_{\phi(q_m)\times \phi(q_2)} & \dots & \frac{1}{\sqrt{N\phi(q_m)}}\bold{I}_{\phi(q_m)} \\
  \end{bmatrix}
\label{QMatrix}.\\
\end{equation}\normalsize
Here $\bold{I}_q$ is an identity matrix of order $q$.
Then the demodulation matrix $\bold{E}_{r}$ is given by,
\begin{equation}
\label{RPTdemod}
\bold{E}_{r} = \begin{cases}{(\bold{Q}\bold{E}_{t}})^{-1}& \text{for $\log_2{N} \notin \mathbb{Z}$}\\
({\bold{Q}\bold{E}_{t}})^{T} & \text{for $\log_2{N} \in \mathbb{Z}$},
\end{cases}
\end{equation}
and $\bold{E}_{r}$  is also known as normalized Ramanujan Periodic Transformation (RPT) matrix \cite{6839030}.
\begin{figure}[ht!]
\centering
\subfloat[][]{
\includegraphics[width=4.5cm,height=3.5cm]{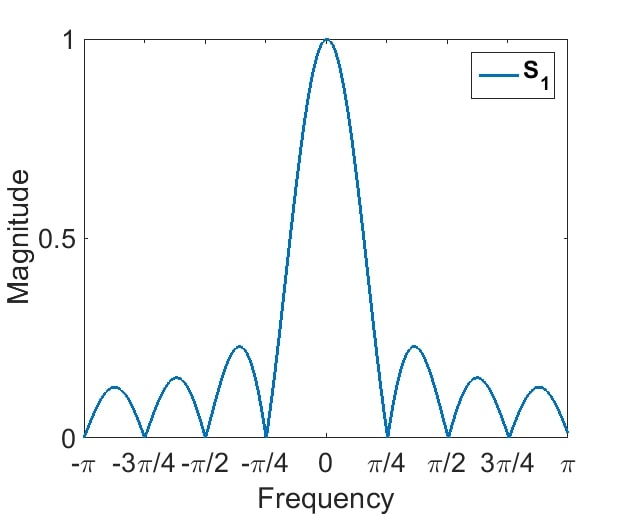}
}
\subfloat[][]{
\includegraphics[width=4.5cm,height=3.5cm]{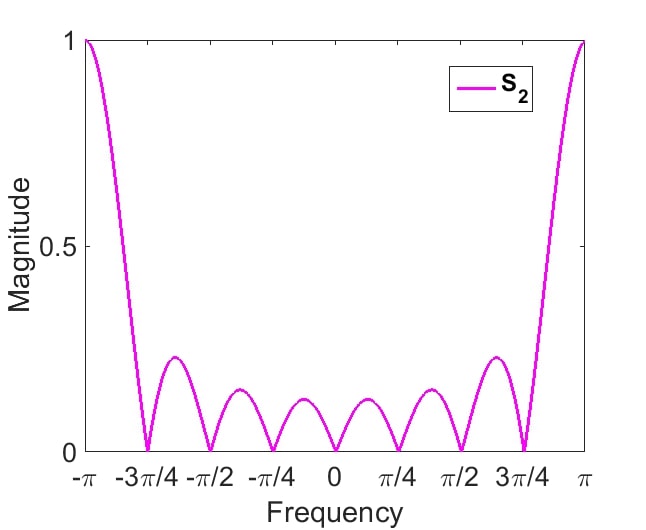}
}\\
\subfloat[][]{
\includegraphics[width=4.5cm,height=3.7cm]{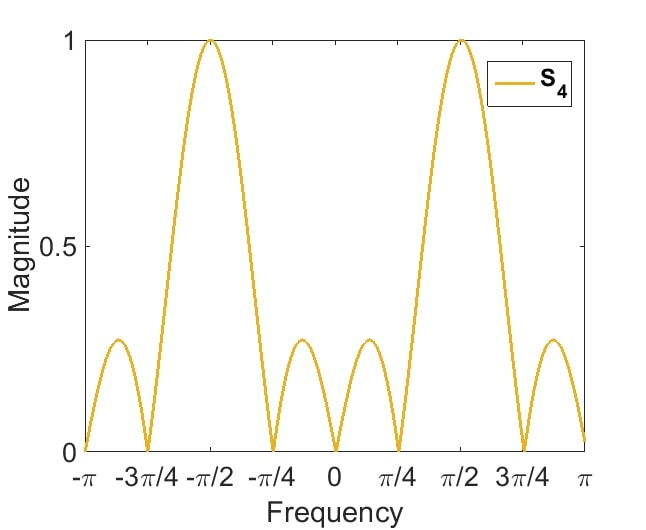}
}
\subfloat[][]{
\includegraphics[width=4.5cm,height=3.7cm]{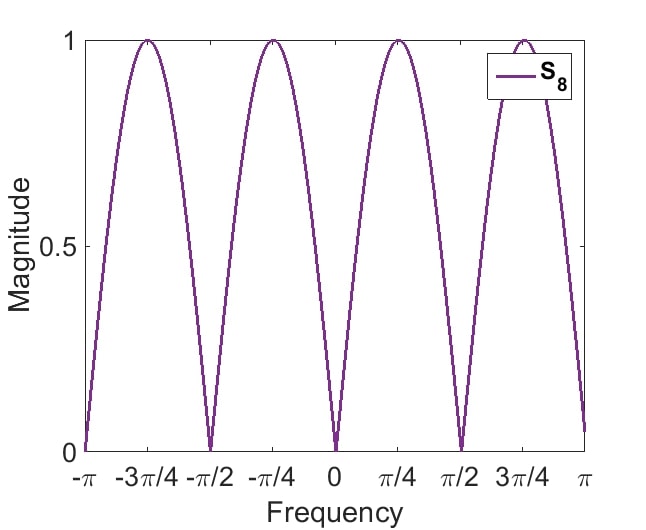}
}
\caption{\small{Ramanujan Subspace Spectrum for N=8, (a) $\bold{S}_1$ (b) $\bold{S}_2$  (c) $\bold{S}_4$  (d) $\bold{S}_8$ }}
\label{fig2}
\end{figure}
\newtheorem{prop}{Proposition}
\begin{prop}
For the LTI multipath fading scenario described in \eqref{sysmod3},   channel matrix $\bold{H}_{cir}$ of size $N \times N$ is transformed to Toeplitz stair block diagonal $\bf{H}_{sbd}$ under pre and post linear transformation $\bold{E}_{t}$ and  $\bold{E}_{r}$ respectively.

\begin{equation}
\bold{H}_{sbd} = \bold{E}_{r}\bold{H}_{cir}\bold{E}_{t},
\label{RPTdecomp}
\end{equation}
\begin{equation}                                                                                                                                                                                                                                                                                                                                                                                                                                                                                                                                                                                                                                                                                                                                                                                                                                                                                                                                                                                                                                                                                                                                                                                                                                                                                                                                                                                                                                                                                                                                                                                                                                                                                                                                                                                                                                                                                                                                                                                                                                                                                                                                                                                                                                                                                                                                                                                                                                                                                                                                                                                                                                                                                                                                                                                                                                                                                                                                                                                                                                                                                                                                                                                                                                                                                                                                                                                                                                                                                                                                                                                                                                                                                                                                                                                                                                                                                                                                                                                                                                                                                                                                                                                                                                                 \bold{H}_{sbd} =\begin{bmatrix}
    \bold{H}_{q_1}& \bold{0}_{\phi(q_1)\times \phi(q_2)} &\dots&\bold{0}_{\phi(q_1)\times \phi(q_m)}\\
    \bold{0}_{\phi(q_2)\times \phi(q_1)}& \bold{H}_{q_2}  & \dots&\bold{0}_{\phi(q_2)\times \phi(q_m)}\\
     \vdots &\vdots & \ddots &\vdots\\
    \bold{0}_{\phi(q_m)\times \phi(q_1)} & \bold{0}_{\phi(q_m)\times \phi(q_2)} & \dots & \bold{H}_{q_m} \\
  \end{bmatrix}
\label{sbd}\\
\end{equation}
\begin{enumerate}
\item  $\bold{H}_{sbd}$ is the stair block diagonal matrix, where the $\bold{H}_{q_i}$ for $q_i \geq 3$ are toeplitz matrices. The number of main block diagonals matrices depends up on '$m$' divisors of $N$ as shown in (\ref{sbd}) and its respective sizes are related to totient  function $\phi(q_i)$.

\item  If $N$ is power of $2$ , then each $\bold{H}_{q_i}$ block results in skew-circulant matrix.

\end{enumerate}
\end{prop}
\emph{Proof:} Provided in appendix I.\\
Therefore channel $\bold{H}$ in \eqref{sysmod_MC} is modified to stair block diagonal matrix $\bf{H}_{sbd}$ under the linear transformations of $\bold{E}_t$ and $\bold{E}_r$,
\begin{equation}
\bold{Y} =  \sqrt{\frac{P}{N}}\bold{H}_{sbd}\bold{X}+\bold{W}.
\label{sysmod_MCR}
\end{equation}
In specific, if $N$ is some integer power of $2$, $\bold{H}_{q_2} \dots \bold{H}_{q_m}$ results in skew-circulant matrices. For instance consider $N=4$, then

\begin{equation}                                                                                                                                                                                                                                                                                                                                                                                                                                                                                                                                                                                                                                                                                                                                                                                                                                                                                                                                                                                                                                                                                                                                                                                                                                                                                                                                                                                                                                                                                                                                                                                                                                                                                                                                                                                                                                                                                                                                                                                                                                                                                                                                                                                                                                                                                                                                                                                                                                                                                                                                                                                                                                                                                                                                                                                                                                                                                                                                                                                                                                                                                                                                                                                                                                                                                                                                                                                                                                                                                                                                                                                                                                                                                                                                                                                                                                                                                                                                                                                                                                                                                                                                                                                                                                               
 \bold{E}_{t} =\setlength\arraycolsep{1pt}
  \begin{bmatrix}
     1 &  \phantom{-}1& \phantom{-}2 & \phantom{-}0\\
     1 &  -1 & \phantom{-}0 & \phantom{-}2\\
      1 &  \phantom{-}1 &-2 & \phantom{-}0\\
       1 & -1 & \phantom{-}0 & -2
  \end{bmatrix}
\label{RPT4}
\end{equation}

\begin{equation}                                                                                                                                                                                                                                                                                                                                                                                                                                                                                                                                                                                                                                                                                                                                                                                                                                                                                                                                                                                                                                                                                                                                                                                                                                                                                                                                                                                                                                                                                                                                                                                                                                                                                                                                                                                                                                                                                                                                                                                                                                                                                                                                                                                                                                                                                                                                                                                                                                                                                                                                                                                                                                                                                                                                                                                                                                                                                                                                                                                                                                                                                                                                                                                                                                                                                                                                                                                                                                                                                                                                                                                                                                                                                                                                                                                                                                                                                                                                                                                                                                                                                                                                                                                                                                               
  \bold{H}_{cir} = \setlength\arraycolsep{2pt}
  \begin{bmatrix}
     -2\matplus4i &  \phantom{-}0\matminus4i & \phantom{-}1 \matminus5i &  \phantom{-}3\matplus0i\\
     \phantom{-}3\matplus0i &  -2\matplus4i& \phantom{-}0 \matminus4i & \phantom{-}1 \matminus5i \\
      \phantom{-}1\matminus5i &\phantom{-} 3\matplus0i & -2\matplus4i&\phantom{-} 0 \matminus4i\\
       \phantom{-}0\matminus4i &  \phantom{-}1\matminus5i & \phantom{-}3\matplus0i & -2\matplus4i
  \end{bmatrix}
\label{circulant4}
\end{equation}

\begin{equation}                                                                                                                                                                                                                                                                                                                                                                                                                                                                                                                                                                                                                                                                                                                                                                                                                                                                                                                                                                                                                                                                                                                                                                                                                                                                                                                                                                                                                                                                                                                                                                                                                                                                                                                                                                                                                                                                                                                                                                                                                                                                                                                                                                                                                                                                                                                                                                                                                                                                                                                                                                                                                                                                                                                                                                                                                                                                                                                                                                                                                                                                                                                                                                                                                                                                                                                                                                                                                                                                                                                                                                                                                                                                                                                                                                                                                                                                                                                                                                                                                                                                                                                                                                                                                                               
\bold{H}_{sbd} =\setlength\arraycolsep{2pt}
  \begin{bmatrix}
     8\matminus20i&  0& 0 &0\\
        0 &  \matminus16\matplus12i& 0 & 0 \\
        0 & 0 & \matminus24\matplus72i& \matminus24\matminus32i\\
       0 & 0 & 24 \matplus 32i &  \matminus24\matplus72i
  \end{bmatrix}.
\label{vbd4}
\end{equation}

\subsection{Detectors}

Assuming known CIR and $\sigma^2$,  $\bold{G}_{r}$ matrix represents ZF and MMSE detectors as  defined below,
\begin{equation}
\label{RPTdemod}
\bold{G}_{r} = \begin{cases}{(\bold{H}^{H}_{d}\bold{H}_d+ \zeta \bold{I}_{N})}^{-1}\bold{H}_d^{H}& \text{for OFDM}\\
{(\bold{H}^{H}_{sbd}\bold{H}_{sbd}+\zeta \bold{I}_{N})}^{-1}\bold{H}_{sbd}^{H}& \text{for RPSDM},
\end{cases}
\end{equation}
where  $\zeta = 0$ for ZF and  $\zeta = \sigma^2$ for MMSE. This $\bold{G}_{r}$ is applied at the receiver with subcarrier wise for OFDM and subspace wise (block) for RPSDM \cite{1425749,8044294}.

\section{PAPR Analysis}\label{PAPRanalysis} 
PAPR for any signal $x[n]$ is defined as the ratio of the maximum instantaneous power to its average power \cite{cho2010mimo}. It is denoted as $\chi$,
\begin{equation}
\chi =  \frac{\substack{\text{max}\\{0\leq n \leq N-1}}\{|x[n]|^{2}\}}{\mathbb{E}\{|x[n]|^{2}\}},
\end{equation}
where, $\mathbb{E}[.]$ is expectation operator. $\chi$ in decibels (dB) is given by,
\begin{equation}
\chi_{dB}  =  10\log_{10}(\chi).
\end{equation}
For a given $M-$QAM constellation and $N$ length transformation, the worst case PAPR \cite{tse2005fundamentals} (It is the ratio of  any farthest signal point power from the origin in the constellation diagram to the average power of the signal points) of OFDM  \cite{5963448} and RPSDM are computed using standard synthesis equation \eqref{ofdmsynth} and \eqref{ramsEq} respectively. Therefore the PAPR of OFDM is represented as,

\begin{equation}\label{PAPR_OFDM}
\text{PAPR-OFDM  }(\chi_{\tiny{OFDM}}) =  \frac{\beta^{2}N}{\alpha^2},
\end{equation}
where, $\alpha^{2}$ and $\beta^{2}N$ are the average power and worst case peak power of an OFDM signal respectively provided in appendix II.

In similar way, worst case PAPR of RPSDM is derived in appendix II and final result is shown in below,
\begin{equation}\label{PAPR_RPSDM}
\text{PAPR-RPSDM  }(\chi_{\tiny{RPSDM}}) =   \frac{\beta^{2}\bigg(\sum_{{q_i}|N}\frac{\gamma_{q_i}}{\sqrt{\phi{(q_i)}}}\bigg)^{2}}{N\alpha^{2}},
\end{equation}
where $\gamma_{q_i} = \sum_{l=0}^{\phi{(q_i)}-1}\hat{c}_{q_i}(l)$.

\section{Results and Comparison}\label{Results}
\begin{figure}
\centering
\includegraphics[width=9cm,height=6cm]{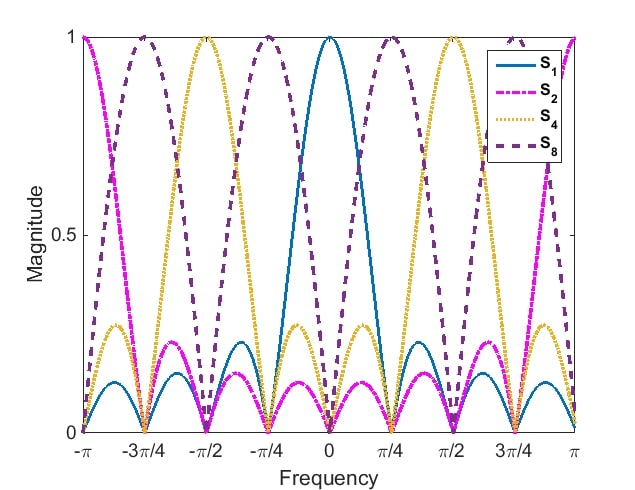}
\caption{\small{Combined Spectrum for $N = 8$ ($\bold{S}_1$,$\bold{S}_2$,$\bold{S}_4$  and $\bold{S}_8)$}}
\label{fig211}
\end{figure}

For a given $N=8$ magnitude frequency spectrum is computed for the basis of each divisor Ramanujan subspace, which are depicted in Fig.~\ref{fig2} (a)-(d). The number of subcarriers present in each subspace $\mathbf{S}_{q_i}$ is defined by its totient function $\phi{(q_i)}$ of that subspace. It is evident from Fig.~\ref{fig2} (a)-(b) that the subspace $\mathbf{S}_1$ and  $\mathbf{S}_2$ has only one subcarrier. The magnitude frequency spectrum of the subspace $\mathbf{S}_4$ spanned by the two basis vectors $c_{4}[n]$ and $c_{4}[n-1]$ are same, which are depicted in Fig.~\ref{fig2} (c). In similar way the magnitude spectrum of $\mathbf{S}_8$ spanned by $c_8[n]$ and its consecutive $3$ circular shifts are also same as depicted in Fig.~\ref{fig2} (d). Fig.~\ref{fig211} represents the non-overlapping magnitude spectrum of each divisors Ramanujan subspace, the same is stated in [section 3.2]. This transform domain visualization of these subspaces may allow variable frequency diversity benefits for transmitting data symbols. 

\begin{figure}[!ht]
\centering
\includegraphics[width=8.8cm,height=6.2cm]{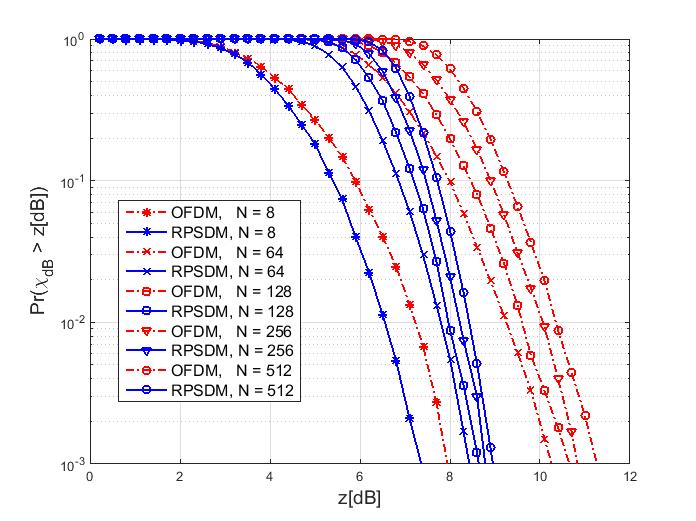}
\caption{\small{CCDF of OFDM  \& RPSDM for N = 8, 64,128, 256 \& 512.}}
\label{fig3}
\end{figure}
\subsection{PAPR Simulation Results}
Fig. \ref{fig3} shows comparison of PAPR performances of OFDM and RPSDM for $N = 8, 64,128, 256$ and $512$ using Complimentary Cumulative Distribution Function (CCDF) as a performance metric. In case of $N=128$, the PAPR values of OFDM and RPSDM for CCDF probability of $10^{-3}$ are 10.5dB and 8.5dB respectively. PAPR performance of RPSDM is lower by 2dB than the OFDM. Table \ref{worstcasePAPR} tabulates the worst case PAPR performance of OFDM \eqref{PAPR_OFDM}  and RPSDM \eqref{PAPR_RPSDM} respectively. 
\begin{table}[ht]
\centering
\begin{tabular}{ | M{5em} |  M{2em} |  M{2em} | M{2em} | M{2em} | M{2em} | M{2em} | M{2em} | }
\hline
Subcarriers & 8&16&32&64&128&256&512\\
 \hline
OFDM \eqref{PAPR_OFDM} & 11.5 & 14.5 &17.60 & 20.61&23.62 & 26.63&29.64\\
 \hline
RPSDM\eqref{PAPR_RPSDM} & 8.19 &8.83 & 9.25& 9.50&9.74 &9.88 & 9.98 \\
\hline
\end{tabular}
\caption{Worst case comparison  of PAPR in (dB)}
\label{worstcasePAPR}
\end{table}
As it is evident from \eqref{PAPR_OFDM} that if $N$ is some integer power of $2$ ($2^{m_p} = 8,16, \dots 512$)  then PAPR increases by 3dB for every increment of $m_p$ by 1. Whereas worst case PAPR of RPSDM is not growing exactly at the same rate as that of OFDM. It is seen from \eqref{PAPR_RPSDM} for every increment of $m_p$ worst case PAPR rate is decreasing. The reason behind this decreasing rate is due to increasing sparsity of first row vector of $\bold{E}_{t}$ used for computing $x[0]$ \eqref{RPSDMpeakpower}.
\begin{figure}[!ht]
\begin{flushleft}
\includegraphics[width=9cm,height=6cm]{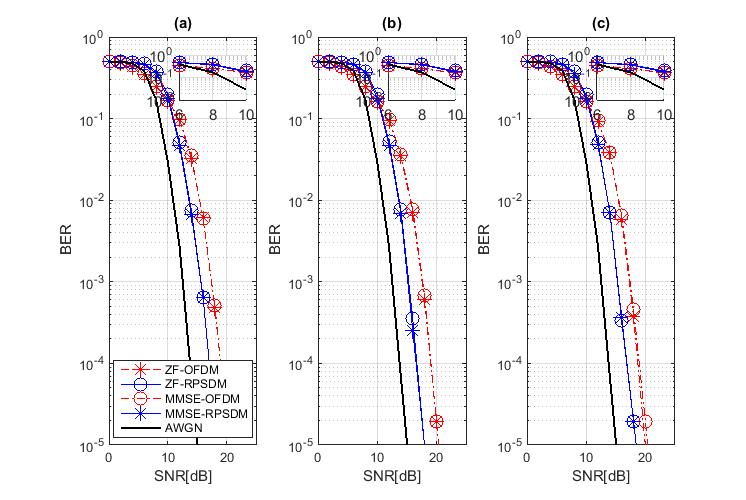}
\caption{\small{BER of ZF and MMSE detector for OFDM and RPSDM (a) N = 128, (b) N = 256 and (c) N = 512.}}
\label{fig5}
\end{flushleft}
\end{figure}
\subsection{BER Simulation Results}
Consider a SISO system with $16$-QAM modulation symbols as input data for BER simulation results. The same system is realized for selective fading channel with multipaths $L=8$ whose coefficients are complex Gaussian random variables with zero mean and unit variance. 
 
Fig. \ref{fig5} shows BER performance of OFDM and RPSDM multicarrier methods along with ZF and MMSE detection for length $N = 128,256$ and $512$. It is seen that for all cases, the RPSDM performs better than OFDM at high SNR and OFDM is better than RPSDM at low SNR. It is concluded that the RPSDM-ZF detector has low noise amplification compared to OFDM-ZF detector \cite{cho2010mimo} at high SNRs (>15dB).  It is also seen that the RPSDM-MMSE detector outperforms the OFDM-MMSE. MMSE detector is slightly better compared to ZF detector in all scenarios. As the MMSE requires noise variance $\sigma^2$ estimate, which results in extra computational complexity. Therefore the ZF with RPSDM detector is chosen over MMSE detector.

\subsection{Computational Complexity}

\begin{figure}[!ht]
\centering
\includegraphics[width=8.5cm,height=6cm]{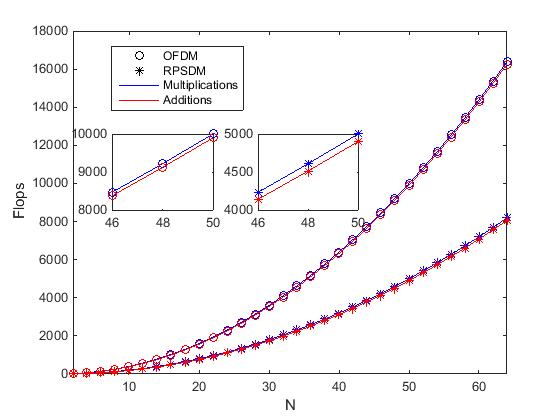}
\caption{\small{Computational Complexity of OFDM and RPSDM.}}
\label{fig31}
\end{figure}
In this subsection, for a given input $x \in \mathbb{C}^{N}$, we compare the complexity analysis of RPSDM and OFDM using direct method (i.e.,  using $\mathbf{E}_{t}$ and $\mathbf{E}_{r}$ ) and fast algorithm based method, particularly for $N$, is some integer power of 2.\\

\emph{Direct Method:} For any arbitrary $N$, the total number of real multiplications ($\mathcal{C}_{m}^{o}$) and additions ($\mathcal{C}_{a}^{o}$) for generating OFDM modulated symbols are \cite{oppenheim2014discrete},
\begin{equation}
\label{OFDMMatrixComplexity}
\mathcal{C}_{m}^{o} = 4N^2, \ \mathcal{C}_{a}^{o} = 2N(2N-1).
\end{equation} 
In a similar way, RPSDM requires $\mathcal{C}_{m}^{r}$  and $\mathcal{C}_{a}^{r}$ integer multiplications  and additions respectively, where,
\begin{equation}
\label{RPSDMMatrixComplexity}
\mathcal{C}_{m}^{r} = 2N^2, \ \mathcal{C}_{a}^{r} = 2N(N-1). 
\end{equation}
From \eqref{OFDMMatrixComplexity} and  \eqref{RPSDMMatrixComplexity}, there is a reduction of 2$N^2$  number of multiplications as well as additions. It is clear that RPSDM saves 50\% computational flops over an OFDM. This reduction is due to integer elements of the RPT matrix. Fig. \ref{fig31} depicts the computational complexity of RPSDM over an OFDM for different $N$ values.\\

\emph{Fast Algorithm Based Method:} For complexity analysis, DIT radix-2 FFT algorithm is used to compute DFT/IDFT. This algorithm requires complex multiplications ($\mathcal{C}_{m}^{of}$) and additions ($\mathcal{C}_{a}^{of}$) are \cite{oppenheim2014discrete},
\begin{equation}
\label{OFDMFastComplexity}
\mathcal{C}_{m}^{of} = \frac{N}{2}\log_2(N), \  \mathcal{C}_{a}^{of} = N\log_2(N),
\end{equation}
This leads to the following number of real multiplications ($\mathcal{C}_{m}^{oft}$) and additions ($\mathcal{C}_{a}^{oft}$) \cite{winograd1978computing},
\begin{equation}
\label{OFDMTotComplexity}
\mathcal{C}_{m}^{oft} = 2N\log_2(N),  \ \mathcal{C}_{a}^{oft} = 3N\log_2(N).
\end{equation}

As best of our knowledge, there is no such fast algorithm based method is available in the literature for computation of RPT coefficients. Therefore from observations, we provide an alternative solution to RPSDM.\\

 \emph{It is noticed that RPT ($\mathbf{E}_{t}$) matrix results in sparse nature for a given $N = 2^{m_p}, \ m_p \in \mathbb{N}$}. The number of non-zero elements in each row of $\mathbf{E}_{t}$ \eqref{RPT4} is $\tau(N)$. In general,
\begin{equation}
\tau(N)  = \substack{\prod \\p_{prime}}(m_p+1),\  m_p \in \mathbb{N},
\end{equation}
is a divisor function for a given $N = \substack{\prod \\ p_{prime}}p^{m_{p}} $ \cite{1407923}. For this scenario  $(N = 2^{m_p})$, $\tau(N) = m_p + 1 = \log_2(N)+1$.\\

Then the complex multiplications ( $\mathcal{C}_{m}^{rs}$) and additions ($\mathcal{C}_{a}^{rs}$) required for Inverse RPT are,
 \begin{equation}
\label{RPSDMSparseComplexity}
\mathcal{C}_{m}^{rs} = N(\log_2(N)+1), \  \mathcal{C}_{a}^{rs} = N\log_2(N).
\end{equation}
 
 This leads to number of real multiplications ( $\mathcal{C}_{m}^{rst}$) and additions ($\mathcal{C}_{a}^{rst}$)  as given below,
 \begin{equation}
\label{RPSDMTotalComplexity}
\mathcal{C}_{m}^{rst} = 2N(\log_2(N)+1), \ \mathcal{C}_{a}^{rst} = 2N\log_2(N).
\end{equation}

From \eqref{OFDMTotComplexity} and \eqref{RPSDMTotalComplexity}, it is seen that the RPSDM symbol generation requires $2N$ additional real multiplications and reduction of $N\log_2(N)$ real additions over an OFDM symbol. The required complexity values for few $N$ are tabulated in Table \ref{complexityOR}.\\
\begin{table}[ht]
\centering
\begin{tabular}{ | M{0.5cm} | M{1.8cm}|M{1.4cm}|| M{1.8cm}| M{1.4cm} | } 
\hline
& \multicolumn{2}{|M{3.4cm}|}{\textbf{OFDM}}&  \multicolumn{2}{|M{3.4cm}|}{\textbf{RPSDM}} \\ 
\hline
$N$ & Multiplications {$\mathcal{C}_{m}^{oft}$\eqref{OFDMTotComplexity}} & Additions {$\mathcal{C}_{a}^{oft}$\eqref{OFDMTotComplexity}}& Multiplications {$\mathcal{C}_{m}^{rst}$\eqref{RPSDMTotalComplexity}} & Additions {$\mathcal{C}_{a}^{rst}$\eqref{RPSDMTotalComplexity}}\\ 
\hline
4& 16 & 24 & 24 & 16\\ 
\hline
16& 128 & 192 & 160 & 128\\ 
\hline
64 & 768 & 1152& 896 & 768\\ 
\hline
256 & 4096 & 6144 & 4608 & 4096\\ 
\hline
\end{tabular}
\caption{Computational Comparison of OFDM with FFT/IFFT and RPSDM using \\ sparse nature of RPT/IRPT for $N$  = 4,16,64 and 256}
\label{complexityOR}
\end{table}
 
\textit{Receiver Complexity:} OFDM modulation/demodulation converts the frequency selective channel into $N$ parallel flat fading channels. Therefore, to recover the symbols from these parallel channels, an $N$ one-tap equalizers are needed and which requires $4N$ real multiplications only. But RPSDM converts the same selective channel into parallel fading stair block subchannels having dimensions of respective Ramanujan subspaces. Therefore to recover symbols from these subchannels, block equalizers are needed. These requires $4\sum_{q_i/N}\phi(q_{i})^{2}$ real multiplications and $\sum_{q_i/N}2\phi(q_{i})(2\phi(q_{i})-1)$ real additions. It is clear that RPSDM equalization has high computational complexity compared to an OFDM. As stair block diagonal matrices are Toeplitz structure, one can look for low complexity detectors and left for the future work.

\subsection{Comparison of OFDM and RPSDM}
The merits and demerits of OFDM  and RPSDM with respect to transformation matrices (RPT and DFT) are tabulated in Table \ref{comparisonOR}.  
\begin{table}[ht]
\centering
\begin{tabular}{ | M{4cm} | M{2cm}| M{1.5cm} | } 
\hline
\textbf{Specifications}& \textbf{OFDM} & \textbf{RPSDM} \\ 
\hline
Basis type & Complex & Integer \\ 
\hline
Transformation Matrix & Symmetric & Non-symmetric \\ 
\hline
Generation of transformation matrix $\bold{E}_{t}$  & $\frac{N(N+1)}{2}$  (Complex)  & $\sum\limits_{q_i/N} q_{i}$ (Integer)\\ 
\hline
Channel Coefficients & N & N \\ 
\hline
 PAPR  & High & Low \\ 
\hline
BER (ZF/MMSE detector)  & High & Low \\ 
\hline
Modulator \& Demodulator & High & Low \\ 
\hline
Modulator \& Demodulator (For $N$ power of 2) & (FFT) Low & (Sparse) Moderate\\ \hline
Receiver Complexity & Low & High\\
\hline
\end{tabular}
\caption{Comparison of OFDM and RPSDM}
\label{comparisonOR}
\end{table}

It is evident from \eqref{eq8}, each basis vector of a Ramanujan subspace is a linear combination of all complex exponentials (harmonics) having the same period. This results in a non-sinusoidal shape of the basis vector. The larger deviation from the sinusoidal shape would happen (moving towards flat shaped function-rectangular) when multiple harmonics results in the same period. Due to this, the PAPR of RPSDM is converging as shown in Table  \ref{worstcasePAPR}. While this is not the case for an OFDM.
Due to integer, circular and sparse nature of Ramanujan subspaces, noise amplification for proposed detectors using $\bold{H}_{sbd}$ is low compared to $\bold{H}_d$. This may be one of the reasons for reduction of BER as shown in Fig. \ref{fig5}. Despite these benefits, usage of CP makes RPSDM spectral inefficient as that of OFDM. To improve the spectral efficiency and latency of RPSDM, filtering methods like FBMC are needed. However, RPSDM provides low PAPR and computational efficiency compared to conventional FBMC \cite{6906043, 6293035}. It is seen that the PAPR of FBMC \cite{6906043} for $N = 64$ subcarriers with CCDF probability of $10^{-3}$ is around 16 dB which is much higher than OFDM (around 10 dB), whereas it is 8dB for RPSDM. 

\section{Conclusions \& Future Scope}\label{confut}
In this paper, we developed a Ramanujan subspace based multiplexing design for multi-carrier wireless systems known as RPSDM. It is shown that the decomposition of circulant channels using RPSDM results in Toeplitz stair block diagonal matrix. The PAPR performance of RPSDM is significantly improved as block size of transmitted data increases in the integer power of $2$.  
BER performance of RPSDM with ZF and MMSE detectors outperforms the OFDM at high SNR regime. It is also seen that the BER performance of MMSE detector is slightly superior to the ZF detector in all scenarios with an expense of noise variance estimate. Our future work explores the possibility of utilizing variable frequency diversity provided by RPSDM for wireless frequency selective fading channels.  

\section*{Acknowledgment}
We are also immensely grateful to Dr. Diviyang Rawal, Dr. Santosh Kumar and Dr. Satyanarayan Reddy for their comments on the earlier versions of the manuscript. We also thank all unanimous reviewers for their valuable comments to improve readability of the manuscript.


\bibliographystyle{ieeetr}
\bibliography{RPSDM}

\section*{Appendix I}
\textbf{Proof for proposition 1:} \\
Let us consider the arbitrary  circulant matrix  $\bold{A}$ of size $q_i \times q_i$. For any RS of $q_i$, we have an integer circulant matrix $\bold{D}_{q_i}$ of size $q_i \times q_i$ shown in \eqref{Dqmatrix}. Then
\begin{equation}\label{circproof}
\bold{\bar{A}}=\bold{D}_{q_i}^{T}\bold{A}\bold{D}_{q_i},
\end{equation}
here $\bold{\bar{A}}$ results in circulant matrix using the fact that the circulant matrices are closed under multiplication \cite{davis2012circulant}. 
Now it is known that the rank of $\bold{D}_{q_i}$\eqref{Dqmatrix} is $\phi{(q_i)}$, so the matrix  $\bold{C}_{q_i}$ \eqref{Cqmatrix} consists of $\phi{(q_i)}$ linear independent columns of $\bold{D}_{q_i}$. Therefore,
\begin{equation}\label{toepzproof}
\bold{\tilde{A}}=\bold{C}_{q_i}^{T}\bold{A}\bold{C}_{q_i},
\end{equation}
where $\bold{\tilde{A}}$ is toeplitz matrix or a sub-matrix of $\bold{\bar{A}}$ with first $\phi{(q_i)}$ rows and columns. For any given circulant matrix of size $q_{i}\times q_{i}$ results in toeplitz after removal of last consecutive rows and columns.

Now consider the circulant matrix $\bold{H}_{cir}$ of size $N$ transformed under $\bold{S}_{q_i}$ as shown below,
\begin{equation}
\bold{H}_{q_i}=\bold{S}_{q_i}^{T}\bold{H}_{cir}\bold{S}_{q_i},
\label{Circuprop1}
\end{equation}
Using \eqref{circproof} and \eqref{toepzproof}, $\bold{H}_{q_i}$ results in toeplitz, since the basis $\bold{S}_{q_i}$ of size $N\times \phi{(q_i)}$  contains $\frac{N}{q_i}$  repetitions of the matrix $\bold{C}_{q_i}$.
Accordingly, $\bold{E}_{t}$ and $\bold{E}_{r}^{T}$ which has the basis of $\bold{S}_{q_i} (q_i$ belongs to all $m$ divisors of $N$), results in block diagonal structure $\bold{H}_{sbd}$ \eqref{sbd} with each block size $\phi(q_i)\times\phi(q_i)$.

%
\section*{Appendix II}
we assume the average power $\mathbb{E}\{|X[.]|^{2}\}$ for a given M-QAM is represented as,
\begin{multline}\label{QAMavgpower}
 \mathbb{E}\{|X[.]|^{2}\} = \alpha^{2}\\ = \frac{1}{M}\sum_{n_2=0}^{\sqrt{M}-1}\sum_{n_1=0}^{\sqrt{M}-1}|(2n_1-1-\sqrt {M})+j(2n_2-1-\sqrt{M})|^2
    \end{multline}
    and worst case peak power $\beta^{2}$ of the $M$-QAM is possible for any one of the following symbols,
    \begin{flalign}\label{QAMpeakpower}
     X[.] = \beta = (\pm \sqrt {M}-1)+j(\pm \sqrt{M}-1), \\
     \beta^2 =  2{(\sqrt{M}-1)^2},
     \end{flalign}
\emph{\textbf{Worst case PAPR Analysis of OFDM:}} \\
From \eqref{ofdmsynth}, average power of an OFDM signal is computed as,
\begin{flalign}
\mathbb{E}\{|x[n]|^{2}\} = \mathbb{E}\Bigg\{\Bigg|\frac{1}{\sqrt{N}}\sum_{k=0}^{N-1}{X}[k]{s}_{k}(n)\Bigg|^{2}\Bigg\},\\
 \leq \mathbb{E}\Bigg\{\frac{1}{N}\sum_{k=0}^{N-1}|{X}[k]|^2|{s}_{k}(n)|^{2}\Bigg\},\\
  \leq \frac{1}{N}\sum_{k=0}^{N-1}\mathbb{E}\{|{X}[k]|^2|\}\Big|e^{j2{\pi}kn/{N}}\Big|^{2},
  \end{flalign}
  \text{using} \eqref{QAMavgpower},
  \begin{flalign}\label{OFDMavgpower}
    \mathbb{E}\{|x[n]|^{2}\} \leq \frac{1}{N}\sum_{k=0}^{N-1}\alpha^{2} = \alpha^{2},
\end{flalign}
Worst case peak power of an OFDM is computed for symbol $\beta$ using \eqref{ofdmsynth} at $n = 0$,
\begin{flalign}
x[0] =\frac{1}{\sqrt{N}}\sum_{k=0}^{N-1}{X}[k]{s}_{k}(0),
\end{flalign}
\text{   where $s_{k}[0] = 1$,  $\forall k$ and using \eqref{QAMpeakpower}},
\begin{flalign}
x[0]  = \frac{1}{\sqrt{N}}\sum_{k=0}^{N-1}\beta = \sqrt{N}\beta,
 \end{flalign}
 \text{and its peak power is,}
 \begin{flalign}\label{OFDMpeakpower}
  |x[0]|^2  = N\beta^{2},
  \end{flalign}
Therefore worst case PAPR  of OFDM using \eqref{OFDMavgpower} and \eqref{OFDMpeakpower} is represented as,
 \begin{flalign}
\chi_{OFDM} =\frac{N\beta^{2}}{\alpha^2}.
\end{flalign}
\emph{\textbf{Worst case PAPR Analysis of RPSDM:}} \\
From \eqref{ramsEq} and \eqref{subspaceeq}, average power of RPSDM signal is computed as,
\begin{multline}
\mathbb{E}\{|x[n]|^{2}\} = \\
\mathbb{E}\Bigg\{\Bigg|\frac{1}{\sqrt{N}}\sum_{{q_i}|N}\frac{1}{\sqrt{\phi{(q_i)}}}\sum_{l=0}^{\phi{(q_i)}-1}{X}[l+q_i-\phi(q_{i})]\hat{c}_{q_i}(n-l)\Bigg|^{2}\Bigg\},\\
 \leq \mathbb{E}\Bigg\{\frac{1}{N}\sum_{{q_i}|N}\frac{1}{\phi{(q_i)}}\sum_{l=0}^{\phi{(q_i)}-1}|{X}[l+q_i-\phi(q_{i})]\hat{c}_{q_i}(n-l)|^{2}\Bigg\},\\
 \leq \frac{1}{N}\sum_{{q_i}|N}\frac{1}{\phi{(q_i)}}\sum_{l=0}^{\phi{(q_i)}-l}\mathbb{E}\bigg\{|{X}[l+q_i-\phi(q_{i})]|^{2}\bigg\}\mathbb{E}\big\{|\hat{c}_{q_i}(n-l)|^{2}\big\},
\end{multline}
 \text{using \eqref{QAMavgpower}, we have $\mathbb{E}\bigg\{|{X}[l+q_i-\phi(q_{i})]|^{2}\bigg\} = \alpha^2$},\\
 \begin{multline}
  \mathbb{E}\{|x[n]|^{2}\}  \leq\frac{ \alpha^{2}}{N}\sum_{{q_i}|N}\frac{1}{\phi{(q_i)}}\sum_{l=0}^{\phi{(q_i)}-l}|\hat{c}_{q_i}(n-l)|^{2},
 \end{multline}
since $|\hat{c}_{q_i}(n-l)|^{2} \leq \sum_{\substack{k=1\\(k,q)=1}}^{q}|e^{j2{\pi}k(n-l)/{q}}|^2 = \phi{(q_i)}$,
\begin{multline}\label{RPSDMavgpower}
  \mathbb{E}\{|x[n]|^{2}\}\leq\frac{ \alpha^{2}}{N}\sum_{{q_i}|N}\frac{1}{\phi{(q_i)}}\sum_{l=0}^{\phi{(q_i)}-l}\phi{(q_i)},\\
 \qquad  \leq \frac{ \alpha^{2}}{N}\sum_{{q_i}|N}\frac{1}{\phi(q_{i})}|\phi(q_{i})|^2,\\
\qquad  \leq \frac{ \alpha^{2}}{N}\sum_{{q_i}|N}\phi(q_{i}) =  \alpha^{2}.\\
\end{multline}
Worst case peak power of RPSDM is computed for symbol $\beta$ using \eqref{ramsEq} at $n = 0$,
 \begin{multline}\label{RPSDMpeakpower}
x[0] =\frac{1}{\sqrt{N}}\sum_{{q_i}|N}\frac{1}{\sqrt{\phi{(q_i)}}}\sum_{l=0}^{\phi{(q_i)}-l}{X}[l+q_i-\phi(q_{i})]\hat{c}_{q_i}(0-l),\\
=\frac{1}{\sqrt{N}}\sum_{{q_i}|N}\frac{1}{\sqrt{\phi{(q_i)}}}\sum_{l=0}^{\phi{(q_i)}-1}{X}[l+q_i-\phi(q_{i})]\hat{c}_{q_i}(l),
\\
  = \frac{\beta}{\sqrt{N}} \sum_{{q_i}|N}\frac{1}{\sqrt{\phi{(q_i)}}}\sum_{l=0}^{\phi{(q_i)}-1}\hat{c}_{q_i}(l) = \frac{\beta}{\sqrt{N}} \sum_{{q_i}|N}\frac{1}{\sqrt{\phi{(q_i)}}}{\gamma}_{q_i},
  \end{multline}
\begin{multline}
\text{ where}, \gamma_{q_i} =\sum_{l=0}^{\phi{(q_i)}-1}\hat{c}_{q_i}(l)=\begin{cases}
   {q_i}-\phi{(q_i)}, &\text{if}\ q_i \text{  is prime},\\
   p^{t-1},&\text{if}\ q_i \text{  power of prime}.\\
 \end{cases}
    \end{multline}
    Therefore worst case PAPR  of RPSDM using \eqref{RPSDMavgpower} and \eqref{RPSDMpeakpower} is represented as,
 \begin{flalign}
\chi_{\tiny{RPSDM}} = \frac{\frac{\beta^{2}}{N}\Bigg(\Big|\sum_{{q_i}|N}\frac{\gamma_{q_i}}{\sqrt{\phi{(q_i)}}}\Big|^{2}\bigg)}{\alpha^{2}}.
\end{flalign}

\end{document}